\documentstyle[12pt]{article}
\begin{document} 
\begin{flushright} {OITS 674}\\
July 1999
\end{flushright}
\vspace*{1cm}

\begin{center}  {\large {\bf Parton Branching in Color Mutation
Model}}
\vskip .75cm
{\bf  Rudolph C. Hwa\footnote{E-mail
address: hwa@oregon.uoregon.edu} and  Yuanfang Wu\footnote{Permanent
Address:  Institute of Particle Physics, Hua Zhong Normal University,
Wuhan, China}}

 {\bf Institute of Theoretical Science and Department of
Physics\\ University of Oregon, Eugene, OR 97403-5203}
\end{center}
\vspace*{.5cm}
\begin{abstract}
The soft production problem in hadronic collisions as described in the
eikonal color mutation branching model is improved in the way that the
initial parton distribution is treated.  Furry branching of the partons is
considered as a means of describing the nonperturbative process of parton
reproduction in soft interaction.  The values of all the
moments, $\left< n\right>$ and $C_q$, for $q=2,\cdots,5$, as well as their
energy dependences can be correctly determined by the use of only two
parameters.

\end{abstract}

\section{Introduction}

Recently a color mutation model \cite{1} (hereafter referred to as I)
was proposed for soft production in hadronic collisions.  It is called
ECOMB, which stands for eikonal color mutation branching.  Unlike
the string model, it starts out with the existence of partons as in the
parton model, and describes the evolution process in which a large
color neutral cluster breaks up into smaller and smaller clusters in
successive fission subprocesses.  Its primary achievement is in the
capability of fitting the intermittency data of NA22 \cite{2}, as no
other dynamical model has.  There is, however, one defect in that
version of ECOMB.  The energy dependence of the initial parton
multiplicity in the range $10 < \sqrt{s} < 100$ GeV  is encoded in an 
{\it ad hoc} way, since the emphasis in I has been placed in the movements
of the partons in the rapidity and color spaces as they undergo color
mutation, not in their generation.  In this paper we remove that defect by
focusing on the branching of partons that leads to the initial parton
distribution considered in I.

It should be made clear that there are two main parts in the overall
process of soft production as described by ECOMB.  One is how the
quarks and gluons in the parton model evolve and end up as
hadrons.  That part has been treated in I.  The other part is
concerned with the number of partons at the beginning of the
evolution process and how it increases with the collision energy.  It
is this part that we treat now.

Both parts are nonperturbative, and both involve branching, but in
different senses.  In I the color mutation process leads to the
branching of color neutral clusters that terminates in hadron
formation.  The total number of partons do not change in that
process.  To be described below is the branching of partons as a
result of the collision between incident hadrons.  In that process the
total number of partons increases with energy; it accounts for the
logarithmic increase of the observed average multiplicity with $s$. 
But even at a fixed $s$, the parton number increases with the
evolution parameter in the branching process so that from a limited
number of valence quarks, a large number of gluons and sea quarks
can be generated by the interaction.  It is the $s$ dependence of
that evolution parameter that is incorporated in a natural way with
the parton branching process that results in the $s$ dependence of
the hadron multiplicity.

\section{Parton Branching}

In I it is shown how the eikonal function $\Omega (R)$ forms the
basis for the geometrical-scaling description of the multiplicity
distribution
\begin{eqnarray} 
P_n = \int dR^2 \left(1 - e ^{-2\Omega (R)} \right) {\bf Q}_n (R) 
\quad ,
\label{1}
\end{eqnarray} 
where ${\bf Q}_n (R)$ is the probability of producing $n$ particles at
the scaled impact parameter $R$.  It is further described in I how
${\bf Q}_n (R)$ is determined in an evolution process, denoted
symbolically by ${\cal E}_{m \rightarrow n}$, such that
\begin{eqnarray}
{\bf Q}_n (R)  = {\cal E}_{m \rightarrow n}\left\{{Q}_m (R) 
\right\}
\quad ,
\label{2}
\end{eqnarray} 
in which $m$ partons evolve by color mutation into $n$ hadrons.  In
I the number of partons is denoted by $\nu$, a symbol which we
now reserve for the number of wounded nucleons in a $pA$
collision, a subject of a separate investigation.  The
probability of producing $m$ partons before the evolution, $Q_m(R)$,
at $R$ is related to the probability of having $m$ partons in a
$\mu$-cut Pomeron by
\begin{eqnarray} 
{\it Q}_m (R)  = \sum^{\infty}_{\mu = 1} \pi_{\mu}(R)
B^{\mu}_m/\sum^{\infty}_{\mu = 1} \pi_{\mu}(R)
\quad ,
\label{3}
\end{eqnarray} 
where
\begin{eqnarray}
 \pi_{\mu}(R) = {\left[2 \Omega (R)  \right]^{\mu}  \over  \mu !}\,
e^{-2 \Omega (R)}
\quad .
\label{4}
\end{eqnarray} 
In I, $B^{\mu}_m$ is assumed to be Poissonian with an
$s$-dependent parametrization of the average $\bar{m}$.  We
now improve on this portion of the formulation by incorporating the
branching dynamics of the Furry process.

Our problem is the determination of the number of partons that are
excited by the collision process before they undergo color mutation,
leading eventually to the observed hadrons.  That number, $m$,
must increase with energy and should depend on the impact
parameter $R$.  By the transformation given in (\ref{3}), the
dependence on $R$ is transferred to the dependence on $\mu$,
which denotes the number of cut Pomerons.  A Pomeron represents
a non-planar diagram with vacuum quantum number exchanged
and can be expressed more readily in terms of the constituents of
the incident particles \cite{3}, instead of the impact parameter of the
collision geometry.  Thus $B^{\mu}_m$ is the appropriate quantity to
study when we ask how the constituents of the incident particles
increase in number through successive emissions to arrive at $m$
partons.

The production of partons in hadronic collisions has been a subject of long
standing.  The  parton cascade model (PCM) \cite{3.5} is among the  more
recent attempts, but it is based entirely on perturbative QCD whose
validity for soft processes is known to be unreliable.  Indeed, the
comparisons of the predictions of PCM with the data are all for $\sqrt
s>200$ GeV, whereas the data for soft production only are for $\sqrt
s<100$ GeV.  Thus we reject the use of perturbative QCD for soft interaction.
Nevertheless, we want to consider the proliferation of partons by a
branching dynamics, which is the universal way that the growth of a
population has been successfully described in many fields.  Toward that
end we adopt the Furry process, originally proposed by cosmic-ray showers
\cite{4}, as the mechanism that generates the partons.  Based on the
assumption that the branching process is purely statisitcal, the evolution
equation that a Furry distribution, 
$F^k_j$, satisfies is 
\begin{eqnarray}
{d  \over  dt} F^k_j = (j - 1 )F^k_{j-1} - j  F^k_j \quad,
\label{5}
\end{eqnarray}
which belongs to the generic form of a stochastic branching equation
\cite{5}. It states that the rate of change of  $F^k_j$ is due to the gain from
the emission by any one of the $j-1$ partons at a previous time step
and the loss from the $j$ partons state when one of them emits
one more.  The time variable $t$ is normalized in such a way that
the coupling strength does not appear explicitly in (\ref{5}).  Since
(\ref{5}) does not rely on the smallness of any coupling, it is not
perturbative. The label
$k$ fixes the initial condition in that at $t = 0$ there are $k$
emitters, i.\ e., $F^k_j (t = 0) = \delta_{jk}$.

The solution of (\ref{5}) is \cite{5}
\begin{eqnarray}
F^k_j(w) = {\Gamma (j)  \over  \Gamma (k) \Gamma(j-k+1)} \left( {1 
\over  w}\right)^k  \left(1 -{1 \over  w}\right)^{j-k}\quad ,
\label{6}
\end{eqnarray}
where $w$ is the evolution parameter, $w = e^t$.  From the
generating function of $F^k_j$, it can easily be shown that the
average $j$ is $\bar{j} = wk$.  Thus $F^k_j(w)$ is specified
by 
\begin{eqnarray}
w = \bar{j}/k \quad ,
\label{7}
\end{eqnarray}  
which simply states that the extent of evolution is related to how
much $\bar{j}$ is increased above the initial number $k$.  In
particle physics $w$ is more meaningful than $t$; moreover,
$\bar{j}$ increases logarithmically with energy.

It is of interest to point out that the Furry distributions can be
related to the negative binomial distribution, $P^k_n$, by \cite{5}
\begin{eqnarray}
 F^k_j \left({\bar{j}  \over  k} \right) = P^k_{j-k} 
\left({\bar{j}  \over  k} - 1 \right) \quad .
\label{8}
\end{eqnarray}
The latter distribution has been used extensively to fit the
high-energy data on multiplicity distributions \cite{6}; however, it is
applied directly to the produced hadrons with $k$ used as an
adjustable parameter, which is shown phenomenologically to
decrease with energy.  In contrast, our consideration of the Furry
distribution is for the branching of partons with $k$ having a more
theoretical basis than being a free parameter.

Our immediate task is to relate $k$ to $\mu$, since $F^k_j$ is to be
used in place of $B^{\mu}_m$ in (\ref{3}).  Consider the high-energy
regime where the Pomeron dominates.  A Pomeron is a nonplanar
diagram with the exchange of a cylinder, not a planar ladder.  A
nonplanar diagram is possible only if more than one constituent line
in each incident particle is involved in the exchange of a ladder. 
The simplest case of one-cylinder exchange involves two partons in
each particle, $(a_1, a_2)$ in particle $a$ and $(b_1, b_2)$ in particle
$b$, say.  A ladder between $a_1$ and $b_1$, and another between 
$a_2$ and $b_2$, together form a nonplanar diagram. Cutting the
rungs of these two ladders gives rise to a $\mu = 1$ cut-Pomeron. 
Each half of the cut diagram has four initial partons connected by
two parallel combs of partons.  The $t$-channel cut diagram can be
viewed in the $s$ channel as successive emissions of partons.  Thus
we can identify the process as having been initiated by four partons
(i.\ e., $k = 4$ initial emitters) and ending in $j$ total number of
final partons.  The fluctuation of $j$ around $\bar{j}$ is to be
specified by $F^k_j$, and $\bar{j}$ increases linearly with the
total rapidity opened up by the collision.

When there are more than one cut Pomeron, then the number of
initial emitters must increase accordingly in order that there can be
more cylinders in the nonplanar diagram, one for each additional cut
Pomeron.  For that reason we set 
\begin{eqnarray} 
k = 4 \mu \quad .
\label{9}
\end{eqnarray} 
 At lower energy, say $\sqrt{s}< 20$ GeV, even though
geometrical scaling may still be valid to justify the use of an
$s$-independent $\Omega(R)$ to describe $\sigma_{\rm tot}$ and
$\sigma_{\rm el}$, the reduced phase space limits the multiplicity of
particles produced.  It means that (\ref{9}) needs to be modified at
low energies by a   factor  that suppresses high
$k$ at low $s$, as $\mu$ is summed to high values in (\ref{3}).

To make contact with $B^\mu_m$, it is necessary to relate $j$ to $m$, in
addition to relating $k$ to $\mu$. So far in this section we have referred to
partons without specifying whether they are quarks, antiquarks or gluons.
We now assert that in soft interaction we only have to consider
the Furry branching of  the gluons for the following reasons. (a) There are
far more gluons than sea quarks. (b) The valence quarks are not to be
counted among the partons that are to hadronize in the central region (as
calculated by ECOMB), since they produce the leading particles in the
fragmentation region. (c) The gluons first emitted by the valence quarks
can initiate the Furry branching process. (d) It is implicit in Eq.\ (\ref{5})
that all emissions are identical in character, thus implying  only
one type of coupling, i.e., three-gluon coupling. (e) For $\mu=1$, the
successive gluon emissions from four gluons form two non-planar ladders;
for $\mu=2$ there are eight initial emitters, and so on. Thus $j$ is to be
identified with the number of gluons at the end of Furry branching. Sea
quarks are obtained by the conversion of those gluons to $q\bar q$ pairs.

In I when $m$ partons (denoted by $\nu$ there) are considered for the
evolution of the color mutation process, it has been explicitly specified that
those partons are quarks and antiquarks that form two overlapping
color-neutral initial clusters. Thus before that evolution process begins, all
gluons are to be converted into
$q\bar q$ pairs in a procedure, called the ``saturation of the sea", used  in
the original recombination model \cite{7}.  It is in that way that the
hadronization of gluons is taken into account without the formation of any
glueballs. 
This saturation of the sea is for accounting the partons produced 
in the central region; dressing of the quarks by gluons in the
nonoverlapping region of the hadronic collision leads only to the leading
particles in the fragmentation region, which is not our concern here.  How
quarks turn into hadrons in the central region has been treated in I.  Here
we simply count each gluon as a $q\bar q$ pair, so 
\begin{equation}
m = 2 j \quad .
\label{10}
\end{equation}

Taking together (\ref{9}) and (\ref{10}), we now have 
\begin{equation}
B^\mu_m = \sum_{j,k}\ F^k_j\,(\bar j/k)\ \delta_{k,4\mu}\ \delta_{m,2j}
\quad.
\label{11}
\end{equation}
We use this formula for all $\sqrt s>20$ GeV. At lower $s$ the inability to
produce many particles due to energy conservation (even though the
masses of hadrons are not considered explicitly here or in I) is effected by
cutting off $k$ at 16.  That means that for $10<\sqrt s < 20$ GeV we
replace $\delta_{k,4\mu}$ in (\ref{11}) by
$\delta_{k,4\mu}\,\theta(4-\mu) + \delta_{k,16}\,\theta(\mu-4)$.  Since
$\pi_\mu(R)$ is small for $\mu\geq 4$ at all $R$ (see Fig.\ 1 in I), only a
small part of the calculation is affected by this correction. The important
point is to recognize that Eqs.\ (\ref{1}) - (\ref{4}), supplemented by
(\ref{11}) completely specify the formalism of the problem in a manner
appropriate for Monte Carlo calculation. The simulated value of $m$ is then
used as the initial number of partons (i.e., quarks and antiquarks) that
undergo color mutation and hadronization as described in ECOMB. The only
adjustable parameters to be used are in the $s$ dependence of $\bar j$, as
will be discussed in the next section.

\section{Multiplicity Distribution}

The multiplicity distribution
of the final state hadrons is calculated on the basis of  Eqs.
(\ref{1}) - (\ref{4}) and (\ref{11}). Among them, (\ref{2})
involves a complicated process that determines where those hadrons go in
the phase space, an aspect of the problem that does not concern us here. 
The hadron multiplicity
$n$ is essentially half the parton multiplicity $m$, since only the
recombination of $q$ and $\bar{q}$ into mesons is considered. 
However, the mesons may be resonances, whose decays make $n$
greater than $m/2$.  For that reason ECOMB must be used in (\ref{2}) to
determine $P_n$, except that $B^{\mu}_m$ in I is now replaced by
(\ref{11}).

In Eqs.\ (\ref{1}) - (\ref{4}) there is no explicit dependence on $s$.  The
only place where $s$ enters is in $\bar{j}$ in (\ref{11}), since the extent of
parton branching depends on
$s$.  The phenomenology of multiplicity distribution in the range
$10 < \sqrt{s} < 100$ GeV confirms KNO scaling \cite{8}, which
places a severe constraint on any model on multiparticle
production.  It requires that $\left< n \right>P_n$ plotted against 
$n/\left< n \right>$ be
independent of $s$, even though $\left< n \right>$ itself increases
logarithmically with $s$.  The $s$-independence of the equations
(\ref{1}) - (\ref{4}) by no means guarantees KNO scaling.   Indeed, it has
been a challenge in I to generate $\ell$n\,$s$ increase of $\left< n \right>$,
while keeping the moments
\begin{eqnarray}
C_q = \left<n^q \right>/ \left<n \right>^q
\label{12}
\end{eqnarray} 
roughly constant.  We now meet that challenge here with
$\bar{j}$ being the only $s$-dependent evolution parameter in
the Furry branching.

We use the parametrization
\begin{eqnarray}
\bar{j} = c_0 + c_1 \ell{\rm n} \sqrt{s} \quad,
\label{13}
\end{eqnarray}
where $c_0$ and $c_1$ are our only two free parameters, adjusted
to fit $\left< n \right>$ vs $\ell$n\,$s$.  The result is shown in Fig.\ 1, for
which we have used
\begin{eqnarray}
c_0 = -7 \quad, \quad \quad c_1 = 6 \quad.
\label{14}
\end{eqnarray} 
Evidently, the fit of $\left< n \right>$ is very good. Its success may
not be surprising given two free parameters.  However, what is striking is
that without any further adjustment of any other parameters the calculated
values of $C_q$ turn out to be essentially constant for $\sqrt{s} > 10$
GeV.  Moreover, their values agree remarkably well with the ISR data
\cite{9}, also shown in Fig.\ 1. It should be noted that for the two
lowest-energy points a correction to (\ref{11}) as remarked below that
equation has been used.

Given how well our calculated $C_q$ moments agree with the data, it
should not be a surprise to see the multiplicity distribution exhibiting
excellent KNO scaling.  That is shown in Fig.\ 2 where $\psi(z)=\left< n
\right> P_n$ is plotted against $z=n/\left< n\right>$ for ten different $s$
values.  The curves are essentially indistinguishable from one another.  We
emphasize that this scaling behavior is a highly nontrivial result and is not
guaranteed by our geometrical scaling formalism.

\section{Conclusion}
The part of ECOMB described in I that specifies the initial parton number
$m$ before color mutation is now amended in this paper. We have used
Furry branching to determine the increase of the number of gluons from
$k$ to $j$. By using only two parameters to describe the $s$ dependence of
$\bar j$, we have been able to determine $\left< n\right>$ and $C_q$,
for $q=2,\cdots,5$, all of which agree with the ISR data over the whole
range
$10<\sqrt s<70$ GeV \cite{9}. Together with its capability to fit the
intermittency data \cite{2} as shown in I, the modified version of ECOMB
has achieved in fitting all essential data on soft production in hadronic
collisions.

With the $s$ dependence under control we can now consider the extension
to higher energies where the hard production of minijets is also important
in addition to soft production.  It is a subject that is a natural extension of
the present work.  Furthermore, the generalization to $pA$ and
$AA$ collisions will eventually lead this line of work to the study of
collision processes that will take place at RHIC.  In such processes the
proper treatment of both the soft and the hard production of particles is
essential.

\vspace*{.3cm}

\begin{center}
\subsubsection*{Acknowledgment}
\end{center}

We are grateful to Zhen Cao for helping us with the use of the Monte Carlo
code that he has developed.  This work was supported in part by U.S.
Department of Energy under Grant No. DE-FG03-96ER40972.

\newpage
\begin{center}
\section*{Figure Captions}
\end{center}
\begin{description}
\item[Fig.\ 1]\quad Average charge multiplicity $\left<n\right>_{ch}$ and
the moments $C_q$ as functions of energy. The data are from \cite{9}; the
lines are calculated results.

\item[Fig.\ 2]\quad Calculated KNO distributions for the 10 $\sqrt s$
points in Fig.\ 1. $\psi(z)=\left< n
\right> P_n$ and $z=n/\left< n\right>$.

\end{description}


\newpage

\begin{thebibliography}{000}

\bibitem{1} Z.\ Cao and R.\ C.\ Hwa, hep-ph/9808399, to be published
in Phys.\ Rev.\ D, referred to as I.

\bibitem{2} I.\ V.\ Ajineko {\it et al}.\ (NA 22), Phys.\ Lett.\ B {\bf
222}, 306 (1989); {\bf 235}, 373 (1990).

\bibitem{3} V.\ V.\ Anisovich, M.\ N.\ Kobrinsky, J.\ Nyiri, and
Yu.\ M.\ Shabelski, {\it Quark Model and High Energy Collisions}
(World Scientific, Singapore, 1985).

\bibitem{3.5} K.\ Geiger, {\it Quark-gluon Plasma 2}, edited by R.\ C.\ Hwa,
(World Scientific, Singapore, 1995), and other references cited therein.

\bibitem{4}W.\ H.\ Furry, Phys.\ Rev.\  {\bf 52}, 569 (1937).

\bibitem{5}R.\ C.\ Hwa, {\it Hadronic Multiparticle Production}, edited
by P.\ Carruthers,  (World Scientific, Singapore, 1988), p.\ 556.

\bibitem{6}
See, for example, {\it Multiparticle Dynamics, Festschrift for L\'{e}on
Van Hove and Proceedings}, La Thuile, Italy, 1989, edited by A.\
Giovannini and W.\ Kittel (World Scientific, Singapore, 1990).

\bibitem{7} 
R.C.\ Hwa, Phys.\ Rev.\ D{\bf 22}, 1593 (1980).

\bibitem{8} Z.\ Koba, H.B.\ Nielsen, and P.\ Olesen, Nucl. Phys. {\bf
B40}, 317 (1972).

\bibitem{9}A.\ Breakston {\it et al}, Phys.\ Rev.\ D {\bf 30}, 528
(1984);  G.J.\ Alner {\it et al}.\ (UA5 Collaboration), Phys.\ Lett.\ {\bf
160B}, 199  (1985); M.\ Adamus {\it et al}.\ (EHS/NA22
Collaboration), Z.\ Phys.\  {\bf C 32}, 475 (1986); M.\ Adamus {\it et
al}.\ (EHS/NA22 Collaboration),  Z.\ Phys.\ {\bf C 39}, 311 (1988).

\end{thebibliography}
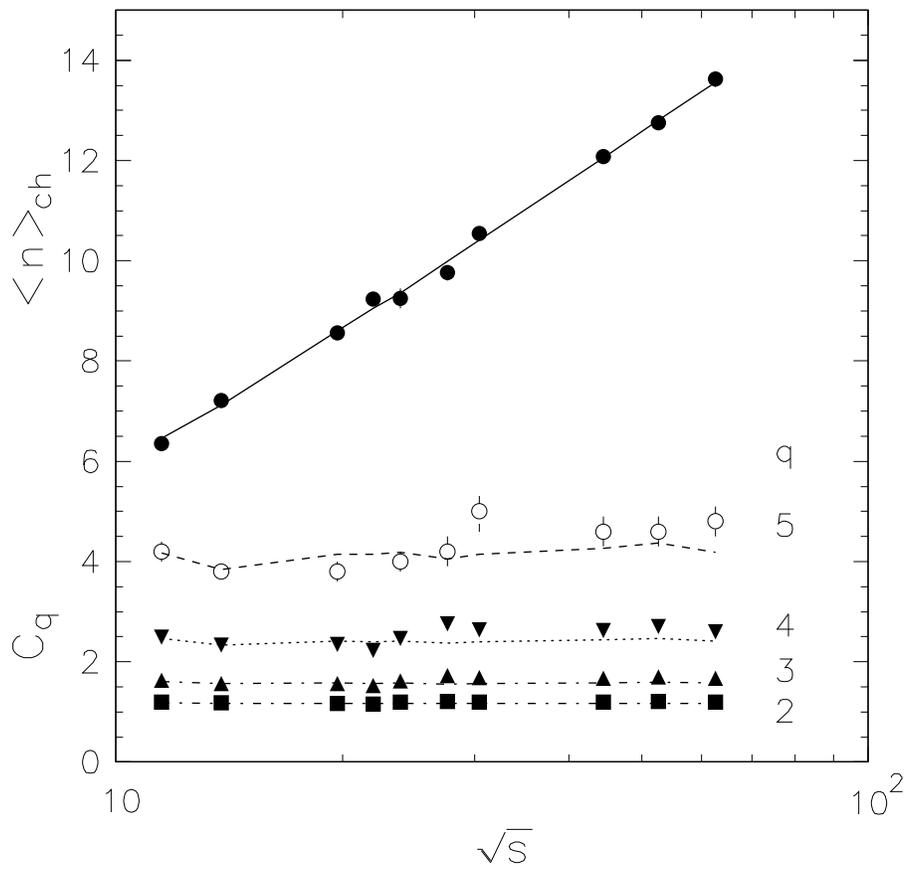
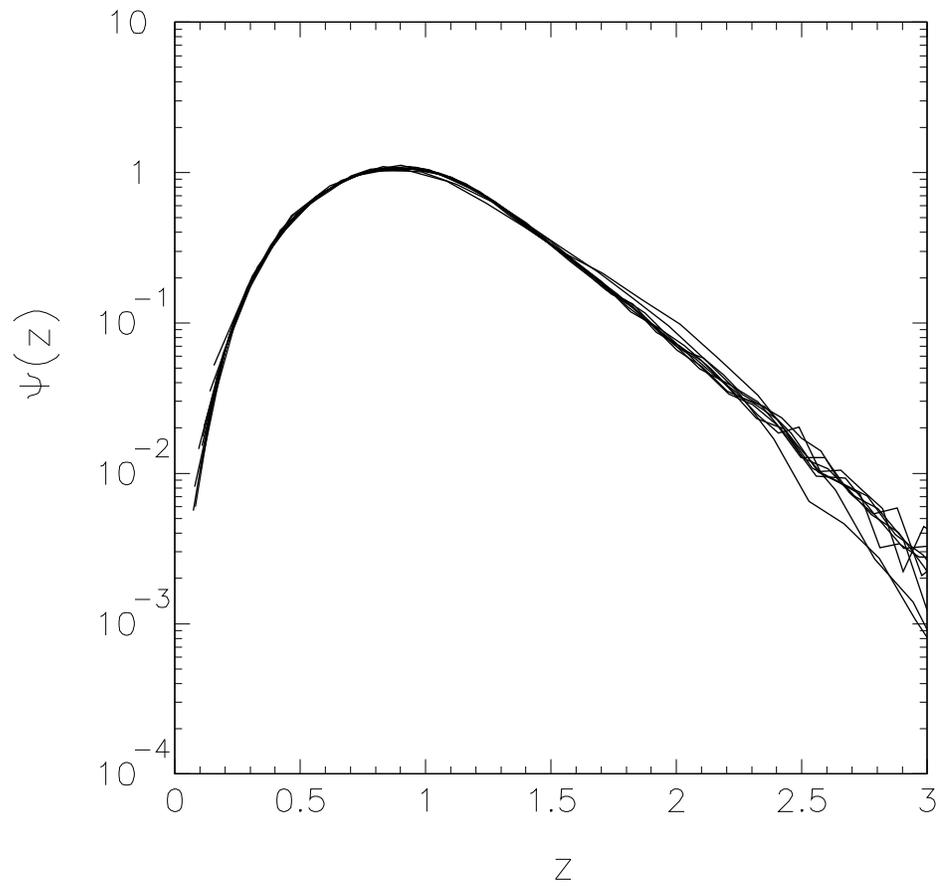
\end{document}